\documentclass[%
 reprint,
superscriptaddress,
 amsmath,amssymb,
 aps,
]{revtex4-2}

\usepackage[english]{babel}

\usepackage{amsmath}
\usepackage{graphicx}
\usepackage[colorlinks=true, allcolors=blue]{hyperref}
\usepackage[autostyle]{csquotes}
\usepackage{multirow}
\usepackage{mathrsfs}
\usepackage{upgreek}
\usepackage{lineno}
\usepackage{float}

\begin{document}

\title{Entropic signature of resonant thermal transport: Ordered form of heat conduction}

\author{Albert Beardo} 
\affiliation{Department of Physics, University of Colorado, Boulder, Colorado 80302}
\affiliation{JILA and STROBE NSF Science and Technology Center, University of Colorado and NIST, Boulder, Colorado 80309}
\author{Prajit Rawte}
\affiliation{Ann and HJ Smead Department of Aerospace Engineering Sciences, University of Colorado Boulder, Boulder, Colorado 80303, USA}
\author{Chia-Nien Tsai}
\affiliation{Ann and HJ Smead Department of Aerospace Engineering Sciences, University of Colorado Boulder, Boulder, Colorado 80303, USA}
\author{Mahmoud I. Hussein}
\affiliation{Department of Physics, University of Colorado, Boulder, Colorado 80302}
\affiliation{Ann and HJ Smead Department of Aerospace Engineering Sciences, University of Colorado Boulder, Boulder, Colorado 80303, USA}


\begin{abstract} 
Thermal transport in crystals is influenced by chemistry, boundaries, and nanostructure. The anharmonic phonon band structure extracted from molecular-dynamics simulations provides an illuminating view of both the type and extent of prevalence of wavelike mechanisms underlying the transport, yet falls short of elucidating the nature of thermal evolution for different phonon regimes. Here we present an analysis framework for the characterization of the entropic signature of the mechanisms induced by boundaries and nanostructure, using both equilibrium and nonequilibrium atomistic simulations. Specifically, we examine the effects of phonon confinement, Bragg scattering, and local resonances on the configurational phase space in room-temperature nanostructured silicon, and quantify how each modifies the~rate of entropy production and thermal relaxation. We reveal that the presence of phonon local resonances spanning the full spectrum enables a highly ordered regime of heat conduction to be approached, where irreversible evolution and entropy maximization are severely hindered by extensive mode hybridizations caused by the resonances. This unique regime of transport paves the way for ultra-precise phonon control for a wide range of applications in condensed matter physics. 
\end{abstract}

\maketitle

\section{Introduction}
The second law of thermodynamics steers the spontaneous evolution of isolated material systems from arbitrary initial conditions towards the maximum entropy state corresponding to equilibrium. In crystal solids, Debye and Peierls, among others, conjectured that the anharmonic nature of atomic interactions combined with the presence of naturally-occurring lattice defects form the key causes behind the tendency of systems to thermally relax~\cite{Peierls1955}.~Consistent with this view, nonlinear phonon-phonon and phonon-defect scattering events are interpreted in the phonon gas picture to not necessarily conserve the lattice quasi-momentum, thus justifying on microscopic grounds the diffusion of vibrational energy along macroscopic thermal gradients. While all atomic interactions and motions are overseen by the Hamiltonian that governs a given system, the specific path and rate of evolution towards thermodynamic equilibrium may be affected by configurational constraints that drive certain mechanistic dynamical behavior.~The inherent competition between mechanical response and the emergence of equilibrium were brought to light by pioneering computational experiments reported by Fermi and colleagues~\cite{Fermi1955}. Lattice disorder and its impact on localization provides an example of a key mechanism that influences the evolution to equilibrium~\cite{anderson1958absence,Lebowitz1974}. More recently, a number of other mechanisms that may hinder thermodynamic behavior have been investigated, which include interaction with thermal reservoirs~\cite{Rieder1967,garcia2017equilibration}, phonon confinement in low-dimensional systems~\cite{balandin1998significant,LEPRI20031}, and nonlinear interfacial effects~\cite{Baowen2004}. \\
\indent Nanostructuring offers yet another rich platform for the disruption of thermal evolution. When feature sizes are confined to the range of the phonon mean free path (MFP) at a given temperature, the intrinsic structural dynamics become a key factor in potentially influencing the path to equilibrium. In this context, a distinct mechanism that has been shown to cause transformative changes to the phonon band structure is atomic-scale resonance hybridizations~\cite{Hussein2014}.~The presence of sub-MFP resonating substructures generates localized phonon resonances that couple with the underlying heat-carrying phonons traveling in the host medium~\cite{hussein2020thermal}. This phenomenon may in principle take place across the full phonon spectrum, thus bringing rise to a unique form of heat conduction that is based on \textit{resonant thermal transport}~\cite{Hussein2018Handbook}. In this regime, spectral energy density (SED) calculations obtained from equilibrium molecular-dynamics (MD) simulations have revealed distinctly appearing flat dispersion curves that couple with the underlying phonon dispersion throughout the full frequency range of atomic motion~\cite{honarvar2016spectral}. Nanostructures exhibiting this behavior, referred to as \textit{nanophononic metamaterials} (NPMs), have recently been experimentally demonstrated~\cite{spann2023semiconductor}. Another wave-based phonon mechanism that also fundamentally influences the phonon band structure is Bragg scattering, which comes into effect with the presence of periodic features smaller than the phonon MFP~\cite{simkin2000minimum,cleland2001thermal,tang2010holey,davis2011thermal}. \\
\indent The impact of the effects mentioned above on the emergence of thermal equilibrium has not yet been studied from a thermodynamics perspective. Yet, in other contexts, analysis techniques and experiments have been developed for the characterization of thermodynamic processes in terms of entropy evolution~\cite{Landi2021}$-$a perspective that in our view opens opportunities for exploring the interplay between structural dynamics and statistical mechanics~\cite{Jarzynski2011,langley2020statistical}. For example, generalized notions of entropy beyond local equilibrium have been applied towards the development of nanoscale thermal transport models not limited to diffusion~\cite{EIT,muller2013rational}. In addition, the equilibrium configurational entropy has been extracted directly from MD simulations~\cite{Baranyai1989,nicholson2021entropy}. Noticeable applications include the characterization of phase transitions~\cite{Piaggi2017}, defects~\cite{Kapur2005,YutianWu2021}, biochemical phenomena~\cite{Xu2023}, and the identification of universal scaling laws of transport properties in liquids~\cite{Dyre2018}. Another key recent advancement is the emergence of thermodynamic uncertainty relations (TURs)~\cite{Pietzonka2017,Horowitz2020} built on fluctuation-dissipation theorems~\cite{seifert2012stochastic} as this provides new tools to relate the fluctuations of dissipative fluxes to the production of entropy in nonequilibrium processes~\cite{Pietzonka2018,Li2019,Saryal2019}.\\

\indent In this article, we quantify$-$for the first time$-$the entropic signature of non-diffusive heat flow in the presence of dynamical constraints created by material nanostructuring, and we do so using MD simulations. The resulting framework bridges the phonon dispersion properties of any given class of nanostructures with the thermodynamic picture and offers a new route to distinguishing between the coherent and noncoherent behavior of phonons.  We consider a variety of crystalline nanoscale features that influence multiple phonon transport mechanisms at room temperature, including boundary scattering, geometric confinement, Bragg scattering, and local resonances. Considering silicon as an example constituent material, for each system we first obtain the phonon band structure by calculating the spectral energy density of atomic motion obtained by MD simulation where anharmonic effects are accounted for. The dispersion relations highlight the influence of the type of nanostructure on the wavelike nature of phonons. Next we characterize the configurational entropy distribution at equilibrium and uncover nonhomogeneous contributions influenced by the presence of the nanoscale features. We then combine nonequilibrium simulations and TUR to obtain a lower bound for the entropy production rate. This metric quantifies the impact of inherent dynamical phenomena on the production and exchange of entropy with the external environment, which allows us to characterize the strength of each configurational constraint in inhibiting the progression towards equilibrium. We find that NPMs are exceptional in this aspect owing to the unique ability of full-spectrum phonon-resonance hybridizations in hindering irreversible evolution.~The existence of this kind of resonant phonon behavior, which manifests as a strongly fluctuating energy current, demonstrates that thermal energy currents do not necessarily correspond to entropy currents at the nanoscale. This brings to light a remarkable observation of a degree of reversibility in phonon transport in the presence of dynamical constraints.

For purposes of comparison, we consider the following geometries: bulk, a suspended membrane, a nanophononic crystal (NPC) with periodic internal cavities, and an NPM in the form of a nanopillared membrane (see geometric details in Fig.~\ref{fig:F1} and Appendix \ref{potential}). We assume the Stillinger-Webber interatomic potential in all models~\cite{SWpotential}. Moreover, the simulated nanostructures are assumed to be defect-free and exhibit atomically flat surfaces. This idealization is made to isolate the aforementioned mechanisms and exclude the effects of impurity scattering and non-pristine phonon-boundary scattering (beyond the effects of atomic relaxation). Future work may examine the effects of material defects and surface roughness consistent with experimental conditions. \\

\section{Anharmonic phonon dispersion relations} 
\begin{figure}
\centering
\includegraphics{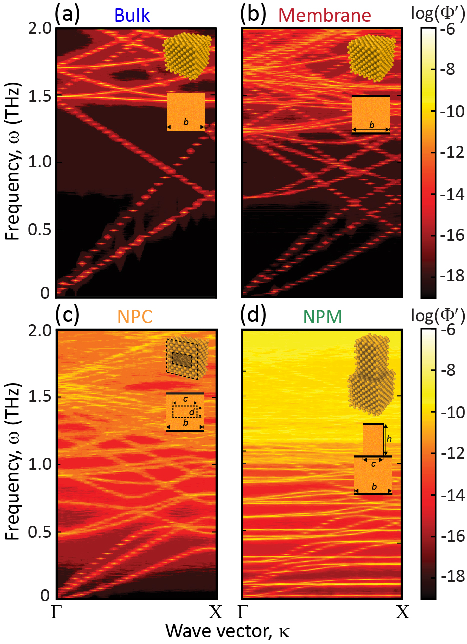}
\caption{\textbf{SED phonon spectrum} of (a) bulk, (b) uniform membrane, (c) a membrane-based NPC, and (d) a membrane-based NPM, all with a membrane thickness of 6 conventional cells (CC). The insets show sketches of the geometry of each unit cell, where $b=6$ CC, $c=4$ CC, $d=2$ CC, $h=18$ CC.}
\label{fig:F1}
\end{figure}

\indent In Fig.~\ref{fig:F1}, we show the phonon spectrum for in-plane wave propagation obtained by applying SED analysis on equilibrium MD simulations~\cite{McGaughey2010,honarvar2016spectral}, as described in Appendix \ref{sed}. Comparing the membrane, NPC, and NPM dispersions with the bulk case, we observe breakings of degeneracies and flattenings of the bands. In the NPM case, in particular, we observe profound changes to the phonon band structure due to the presence of a multitude of local resonances associated with the standing nanopillar~\cite{Hussein2014,wei2015phonon,xiong2016blocking,honarvar2016spectral,Hossein2018,Hussein2018Handbook,hussein2020thermal,spann2023semiconductor}. These resonances, also described as vibrons~\cite{Hossein2018}, appear in the form of horizontal branches at different frequencies spanning the full frequency range of the constituent material, in this case silicon. This feature represents the key character of an NPM. Its significance stems from the couplings with the underlying phonon dispersion curves associated with the base membrane, causing resonance hybridizations that reduce the group velocities of the heat carrying modes and provide a channel for phonon localization. These phonon band-structure features contribute to significant reduction in the in-plane thermal conductivity~\cite{Hossein2018} and recently have been shown to cause distinct phonon localization along the direction of transport within anharmonic MD simulations~\cite{beardo2024resonant}. Also observed are some changes in the linewidths of the membrane, NPC, and NPM dispersion curves compared to the bulk case due to phonon-boundary scattering and lifetime reduction. While the SED spectrum reveals key information on the coherent and incoherent phonon dynamics, it fails to describe the nature of the thermodynamic evolution towards equilibrium. This motivates the following investigation to formally characterize the entropic properties associated with each nanostructure feature.

\section{Equilibrium configurational entropy}\label{Section:configurationalentropy}

First, we study the available phase space of atomic configurations by locally quantifying the configurational entropy at equilibrium. We use an approximate expression derived from an expansion of the configurational entropy in terms of the multibody correlation function~\cite{Baranyai1989}. In liquids, the two-body correlation term is enough to characterize most of the configurational entropy~\cite{Dyre2018}, and it
has also been shown useful to study crystalline solids such
as silicon \cite{Kapur2005,Sluss2022}. This term can be evaluated locally and for a given atom $i$ in the system reads
   \begin{equation}\label{excess_entorpy}
        s^i=-2\pi\rho k_\mathrm{B}\int_0^{r_{m}} [g^i(r)\ln g^i(r)-g^i(r)+1]r^2 \mathrm{d}r,
   \end{equation}
where $\rho$ is the local density around the atom,  $k_\mathrm{B}$ is the Boltzmann constant, $g^i(r)$ is a smoothened radial distribution function, and $r_{m}$ is a cutoff radius~\cite{Piaggi2017}. For comparison purposes, $s^i$ can be interpreted as an estimation of the excess entropy of an atom relative to the ideal gas. Then, the total entropy of atom $i$ can be expressed as $S^i=s^i+S_{\rm IG}$, where the entropy per atom of the ideal gas $S_{\rm IG}$ is evaluated using the Sackur-Tetrode expression. Details on the evaluation of these expressions in equilibrium MD simulations at room temperature are provided in Appendix \ref{eq_MD}, along with a convergence study and verification of the thermodynamic relation between $S$ and the internal energy.

   \begin{figure}
\centering
\includegraphics{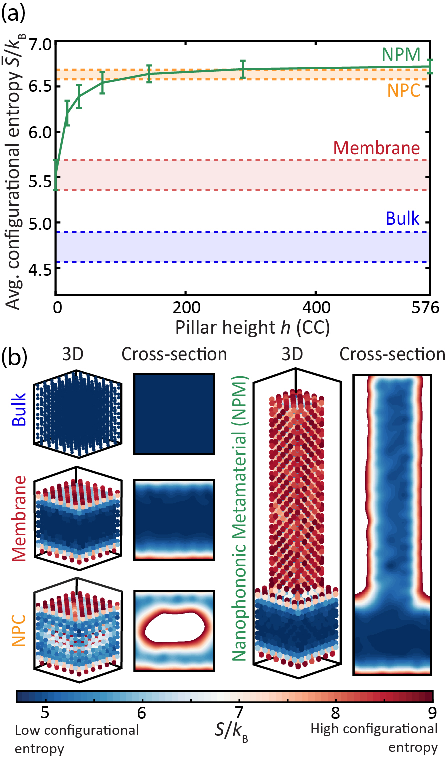}
\caption{\textbf{Equilibrium configurational entropy} average (a) and distribution (b) in the different systems. A relative increase in $S$ is observed close to the free boundaries and in the internal region of the resonating nanopillar.}
\label{fig:F2}
\end{figure}

In Fig.~\ref{fig:F2}, we show local entropy maps for the different systems. Since we use the local density around the atom in Eq.~\eqref{excess_entorpy}, these calculations enable comparison of the available configurational phase space in different regions. We observe a larger accumulation of entropy close to the boundaries compared to the internal regions. Away from the boundaries, the crystalline order is enforced in all directions, thus strongly constraining each atom's dynamics and reducing the size of the phase space of atomic configurations. In contrast, the atoms can accommodate a larger variety of microscopic configurations close to the boundaries due to the reduced number of neighbors. Consequently, the average equilibrium entropy $\bar{S}=\sum_i S^i/N$, where $N$ is the total number of atoms, increases with the surface-to-volume ratio, as clearly shown in the figure, for example, when comparing the membrane and NPC cases with the bulk case. This correlates with the increased deviation of the dispersion from the bulk configuration as observed in the corresponding phonon band structures in Fig.~\ref{fig:F1}. In addition, we observe that the entropy in the interior of the resonating nanopillars is significantly larger than in the interior of the membrane. To quantify this difference, we calculated the average local configurational entropy in the center of the nanopillar (at a distance of 2 CC from its lateral boundaries) and compared it with the average local configurational entropy inside the membrane at a distance of 2 CC from the top and bottom boundaries. We found that the local average in the nanopillar is nearly 25\% greater than that in the membrane. We attribute this to the resonant modes in the nanopillar, which can correlate the dynamics of non-neighboring atoms, thus unlocking microstates that are not accessible in the membrane. Remarkably, this produces a significant increase in the size of the phase space in the NPM. We note that the observed differences in configurational entropy imply that each system needs to produce or exchange different amounts of entropy to reach equilibrium from given nonequilibrium conditions.

\section{Entropy production}\label{Section:entropyproduction}
      
To obtain a comprehensive thermodynamic characterization, we now proceed to study the nonequilibrium evolution of the entropy in each case within the framework of stochastic thermodynamics \cite{Jarzynski2011,seifert2012stochastic}. The evolution of the systems is considered as a sequence of transitions between distinct microscopic configurations mediated by external thermodynamic reservoirs. Due to local detailed balance, a transition in a system from configuration \enquote{x} to configuration \enquote{y} is balanced by a non-dimensional change in entropy of the mediating reservoirs, i.e., 
   \begin{equation}\label{entropy_change}
         \Delta S(\text{x}\rightarrow \text{y})=\ln\frac{r(\text{x}\rightarrow \text{y})}{r(\text{y}\rightarrow \text{x})},
   \end{equation}
where $r$ is the directional transition rate between the two states \cite{Horowitz2020}. While no entropy is generated for reversible processes in which the transition rates are symmetric, asymmetric transitions produce entropy and cause the system to evolve towards the maximum entropy state. 

A microscopic configuration of a crystal lattice can be interpreted in the reciprocal representation in terms of the phonon population, which only maximizes the entropy when accommodating the Bose-Einstein distribution. Therefore, the out-of-equilibrium phonon evolution in the different systems can be investigated in terms of entropy evolution. Here we use steady-state nonequilibrium MD simulations, where the terminals of the system are connected to Langevin thermostats at $T_\text{C}=290$ and $T_\text{H}=310$ K, respectively. To attenuate unphysical coupling/interaction between thermostats, we consider five unit cells between the thermal baths (see Appendix \ref{neq_MD} for details). In steady-state conditions, the entropy that the system produces internally is constantly exchanged with the thermostats, and quantifies the irreversible exchange of energy, or dissipation, to the system surroundings. According to the TUR \cite{Pietzonka2017}, the thermodynamic uncertainty in terms of the energy current established in the system provides a lower bound on the entropy production $\Sigma_\tau$ during a time-window $\tau$:
   \begin{equation}\label{TUR}
         \Sigma_\tau\geq \widehat{\Sigma}_\tau=\frac{2 k_\mathrm{B} \text{E}(q_\tau)^2}{\text{Var}(q_\tau)},
   \end{equation}
where $\text{E}(q_\tau)$ and $\text{Var}(q_\tau)$ are the mean and variance, respectively, of the energy $q_\tau$ exchanged between thermostats through the system during a time-window $\tau$. According to this theorem, an enhancement of the normalized energy current fluctuations indicates a reduction of the minimum entropy production rate required to sustain the steady state. As illustrated by Eq.~(\ref{entropy_change}), such reduction on $\widehat{\Sigma}_\tau$ implies a higher probability of reversible evolution, corresponding to fewer phonon transitions during the time-window $\tau$, and/or, a lower ratio of asymmetric-to-symmetric transitions. Hence, $\widehat{\Sigma}_\tau$ can be used to characterize the influence of system nanostructure on the phonon dynamics and transition rates mediated by reservoirs. Since the wave-particle phonon duality is inherently accounted for, and no assumptions on the nonequilibrium phonon distribution are required, this approach provides insights that are complementary to phenomenological boundary scattering models \cite{mcbennett2023universal} and models based on \textit{ab initio} phonon collision integrals \cite{esfarjani2011heat,seiler2021accessing}.
   
In such a stochastic interpretation of entropy production, the fraction of energy exchange with the thermostats that drive reversible phonon transitions in the system is not identified as energy dissipation and entropy exchange. Consistently, the entropy bound from the present MD simulations is smaller than the entropy exchange in the case of maximum energy dissipation (i.e., $\text{E}(q_\tau)/T_\text{C}-\text{E}(q_\tau)/T_\text{H}>\widehat{\Sigma}_\tau$) for all the cases considered, which in turn validates the TUR as a suitable tool in the present context~\cite{Saryal2019,Supriya2020,Loos2023}. Expression \eqref{TUR} is adequate to study the present systems far from equilibrium (where arbitrary non-Gaussian fluctuations may occur) beyond the linear-response interpretation \cite{Gingrich2016}. Hence, $\widehat{\Sigma}_\tau$ is more fundamental, and robust, than other metrics characterizing nanoscale nonequilibrium conditions such as the thermal conductivity $k$, which is generally a phenomenological concept. Quantification of $k$, using the Green-Kubo method~\cite{zwanzig1965time,kubo2012statistical} or directly by nonequilibrium simulations~\cite{tenenbaum1982stationary,mountain1983thermal}, for example, requires the assumption of Fourier's law$-$which is not necessarily valid at the nanoscale~\cite{benenti2023non}. In contrast, quantification of $\widehat{\Sigma}_\tau$ does not require imposing any constitutive relation between the heat flux and the thermal gradient, which extends the validity of the analysis to complex nanoscale conditions well beyond the diffusive transport regime. Nevertheless, it is illustrative to establish a relation between $\widehat{\Sigma}_\tau$ and an approximate bound for the thermal conductivity $k$ for cases displaying a linear response. If we consider the entropy production rate in terms of the entropy fluxes $\Sigma_\tau/\tau=j/T_\text{C}-j/T_\text{H}$, with $j$ being the energy dissipation rate, and assume Fourier's law, $\widehat{\Sigma}_\tau$ can be proportionally related to a lower bound on the thermal conductivity \cite{Klemens1989}, 
\begin{equation}\label{kappaVSentropy}
    \widehat{k}\eqsim\frac{\widehat{\Sigma}_\tau T^2}{\tau V|\nabla{T}|^2},
\end{equation} 
where $V$ is the system volume, $T$ is the average temperature, and $\nabla T$ is the thermal gradient imposed between the thermostats\footnote{For the bulk case, the thermal conductivity lower bound from Eq.~(\ref{kappaVSentropy}) is approximately five times smaller than what is predicted by comparing the heat flux with the thermal gradient in nonequilibrium simulations.}.

\begin{figure}
\centering
\includegraphics{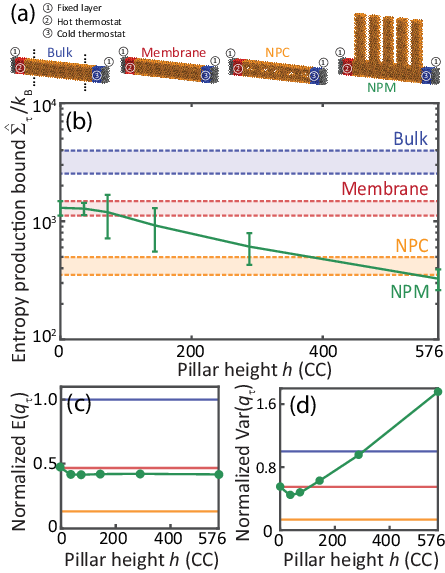}
\caption{\textbf{Nonequilibrium entropy production} for four material systems considered with computational models displayed in (a). Shown are the (b) bound, and normalized (c) mean and (d) variance of the energy transferred between thermostats $q_\tau$ through each material system over a time-window $\tau=$1 ns. In the NPM case, a decrease in the entropy production bound is observed when the nanopillar height is increased$-$a behavior associated with the energy current fluctuations induced by the resonant dynamics of the nanopillars.}
\label{fig:F3}
\end{figure}

In Fig.~\ref{fig:F3}(b), we plot the entropy production bound $\widehat{\Sigma}_\tau/k_\mathrm{B}$ obtained from our nonequilibrium MD simulations considering $\tau=1$ ns. As demonstrated in Appendix \ref{neq_MD}, this is the smallest sampling time for which we obtain $\text{Var}(q_\tau)/\text{E}({q}_\tau)^2\sim 1/\tau$. This relation is a consequence of the central limit theorem in the long sampling time limit \cite{Song2021}, and allows us to characterize a bound for the entropy production rate $\widehat{\Sigma}_\tau/\tau$ independent of $\tau$. Since $q_\tau$ fluctuates both in time and between different simulation realizations, we perform the averages during 30 ns over five different simulations in each case. Analysis of the computational size effects and the influence of the temperature difference between thermostats is provided in Appendix \ref{neq_MD}. As expected, the bulk is the system where entropy is produced most efficiently due to the absence of external constraints other than the thermostats. In contrast, the membrane exhibits a significant reduction in the entropy production rate. This is interpreted in terms of the local constraints to the phonon dynamics close to the free surfaces, which, in combination with the energy conservation restriction, inhibit some asymmetric transitions that are probable in the bulk case, thus reducing the overall probability of irreversible trajectories in the phase space that maximize the entropy. This is consistent with the lower $\widehat{\Sigma}_\tau$ observed in the NPC, where the addition of an internal cavity enhances the phonon-boundary interaction. In this case, however, we note that the entropy production reduction relative to bulk is too large to be solely attributed to more incoherent scattering due to an increase in the surface-to-volume ratio. Instead, the emergence of coherent interference of phonons, i.e., Bragg scattering, represents an additional constraint that can limit the efficient production of entropy. Finally, in the NPM, the addition of the nanopillars is shown to significantly reduce the production of entropy relative to the membrane even though no internal boundaries are included in the basis membrane, where the energy flows between the thermostats. Furthermore, the stronger the resonant effect due to a taller nanopillar, the lower the entropy the system produces. Considering the approximate relation between $\widehat{\Sigma}$ and $\widehat{k}$, this finding suggests that, despite the strong anharmonicity that governs a silicon crystal, introducing local resonances across the full spectrum is an efficient strategy to manipulate nanoscale thermal transport~\cite{Hussein2014,Donadio2020}.   \\
\indent The rate of entropy production also provides a means to quantitatively distinguish between the roles of coherent dynamical effects associated with any existing wave behavior of phonon motion versus noncoherent effects associated with nonlinear phonon interactions and scattering from non-pristine surfaces. In Figs.~\ref{fig:F3}(c) and~\ref{fig:F3}(d), we show $\text{E}(q_\tau)$ and $\text{Var}(q_\tau)$, respectively. Noticeably, boundary scattering in the membrane and the NPC reduces the entropy production mainly by reducing the average energy flux, which also implies a strong variance reduction, while, in the NPM, $\widehat{\Sigma}_\tau$ is reduced mainly due to an increase in flux fluctuations. The strong energy current fluctuations established in the NPM models are thus signatures of the prevalence of coherent phonon effects generated by the nanopillar resonances, while incoherent dynamics dominated by anharmonic and boundary scattering are characterized solely by a reduction in the energy transfer rate. Precisely, in the NPMs we observe that the flux variance increases monotonically by enhancing the resonant effect, even though the average energy current is not significantly modified. According to Eqs.~\eqref{entropy_change} and \eqref{TUR}, this implies that the energy transfer mechanisms in the presence of local resonances involve a reduced amount of irreversible internal phonon transitions. This unique finding offers a broader view that nanoscale structures accommodating strong dynamical constraints can induce some degree of reversible evolution while sustaining thermal energy currents.

\section{Heat flux autocorrelation in the presence of resonances}
The strong current fluctuations in NPMs are associated with resonant hybridizations of propagating phonon modes in the base membrane. In this section, we explore the origin and manifestation of this phenomenon using equilibrium MD simulations. Specifically, we first thermalize the systems following the protocol described in Appendix \ref{eq_MD}. Thereafter and under the resulting equilibrium conditions at room temperature, we analyze the heat flux autocorrelation function (ACF) $<\textbf{J}(0)\otimes \textbf{J}(t)>$, where $t$ is the delay time and $\textbf{J}$ is the heat flux vector integrated over the volume and sampled every 4 fs. For each nanostructure, we perform the simulations in the supercell consisting of five unit cells also considered in the previous section.

\begin{figure*} [t!]
\centering
\includegraphics{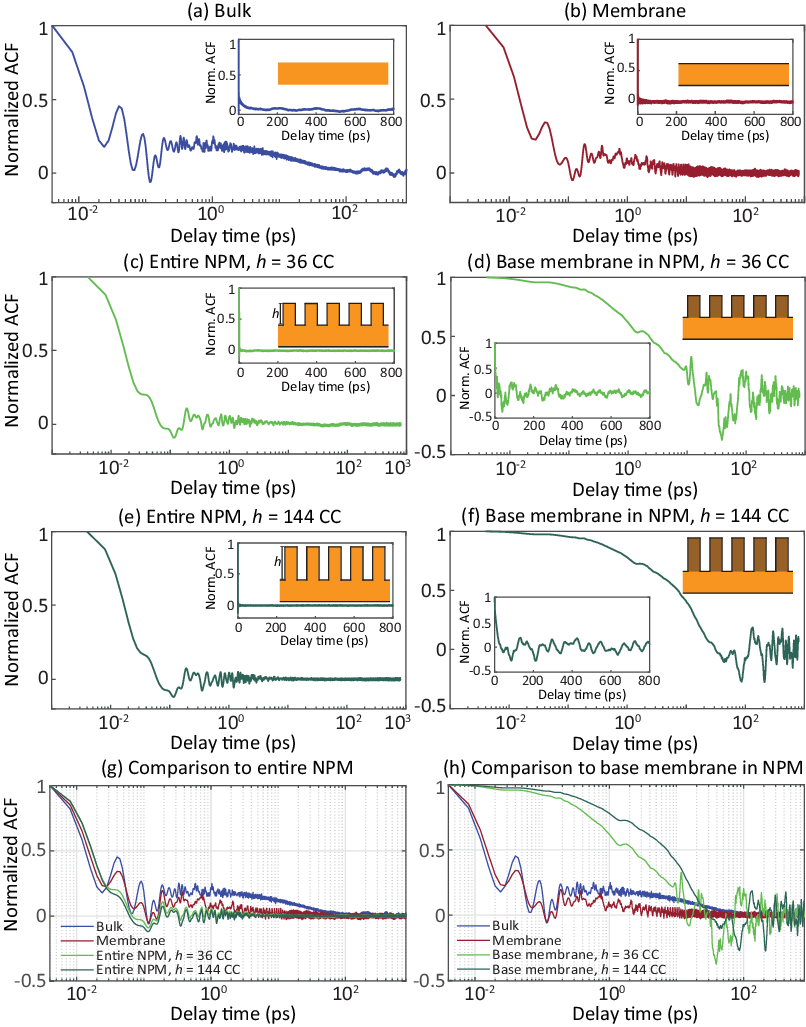}
\caption{\textbf{Heat flux ACF.} The heat flux is extracted from equilibrium simulations at room temperature for (a) the bulk, (b) the membrane and (c-f) the NPM with different nanopillar heights. The heat flux is obtained from the entire simulation domains (c,e) and only from the base-membrane portion (d,f). Direct comparisons between the various ACF functions are illustrated in (g,h). The flux ACF in the base membrane of an NPM displays remarkably slow decay followed by an oscillatory behavior, which indicates the strong coherent coupling between the membrane and the nanopillar.}
\label{fig:Fig_04}
\end{figure*}

In Fig.~\ref{fig:Fig_04}, we show the heat flux ACF for the bulk material, the membrane, and two cases of NPMs with different nanopillar heights $h$. We observe that the autocorrelation of the thermal fluctuations in the NPMs decays faster than in the bulk or the membrane cases, and tends to zero over a time scale of tens of picoseconds. This indicates that the atom dynamics in different regions of the NPM system are not fully correlated, and suggests that thermodynamic equilibrium with ergodic evolution is robustly established in our simulations. Even though our simulation domains are relatively small and sustain strong dynamical constraints, the mixing condition is thus satisfied, which in turn justifies the applicability of the fluctuation theorem and the TUR. 

To examine the influence of resonances in more detail, in Fig.~\ref{fig:Fig_04} we also show the ACF by extracting the heat flux $\textbf{J}$ only from the base-membrane portion of the NPM. The local heat flux ACF displays a remarkably slow initial decay followed by periodic oscillations that persist over a nanosecond time scale. This phenomenon is not observed in the bulk, membrane, or NPC systems, and becomes more pronounced in the presence of taller nanopillars, where the resonant effect is enhanced. Therefore, these features are a manifestation of coherent phonon dynamics in NPMs, and further confirm that the energy transport mechanisms in the base membrane are strongly influenced by the phonon coupling with the resonant modes in the nanopillars. More specifically, the heat flux ACF in the base membrane of an NPM decays slower than in the membrane case. This indicates that the NPM resonances, in fact, work towards increasing the phonon correlations, rather than reducing them as done by scattering from non-pristine boundaries. In analogy with these results, previous work demonstrated that similar features in the heat flux ACF can appear in core-shell nanowires due to the emergence of resonant transversal modes that couple with the propagating longitudinal modes \cite{JieChen2011}$-$a mechanism that is distinctly different from the resonance hybridizations that occur in NPMs, yet still brings rise to this type of behavior.

The anomalous behavior of the heat flux ACF under the influence of atomic-scale phonon resonances at equilibrium is a counterpart of the enhancement of the energy transfer fluctuations in nonequilibrium conditions (see Fig.~\ref{fig:F3}). The coherent coupling of the phonon modes in the nanopillars and the base membrane impairs the stochastic evolution of the vibrational energy distribution, which obstructs the emergence of thermodynamic equilibrium. This suggests that ergodic exploration of the phase space is not as easily accessible in material systems presenting resonances compared to conventional materials, and might indicate a physical origin underlying errors and uncertainty in Green-Kubo calculations~\cite{Kang2017,Souza2017} to extract the thermal conductivity in complex nanoscale configurations prone to various forms of resonances \cite{JieChen2011}. Nevertheless, quantification of the heat flux uncorrelation time and the conductivity was shown possible in NPMs with certain dimensions~\cite{Honarvar2018,HusseinAFM2020} and other complex systems \cite{Giulia2012}.

\section{Entropy maximization time}

The ratio of the equilibrium entropy obtained in Section \ref{Section:configurationalentropy} to the entropy production rate in Section \ref{Section:entropyproduction} is a measure of the entropy maximization time,
\begin{equation}
    \tau_S=\frac{N\bar{S}}{\widehat{\Sigma}_\tau/\tau}.
\end{equation}
This metric characterizes the time required by a system to evolve from the minimum-entropy state, corresponding to a fully coherent and non-thermalized state, towards equilibrium at the given temperature. However, $\tau_S$ does not correspond to actual thermalization times because the production rates $\widehat{\Sigma}_\tau/\tau$ are characterized in steady-state conditions at 300 K, whereas the entropy evolution is a function of temperature and sensitive to the specific nonequilibrium transient trajectory. Nevertheless, this idealized time enables equal-footing comparison of the influence of different dynamical constraints on the emergence of thermal equilibrium.

\begin{figure}[t!]
\centering
\includegraphics{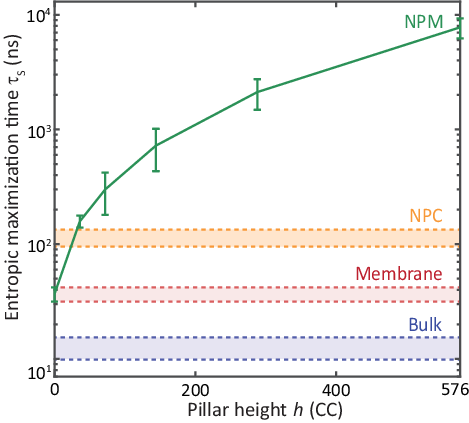}
\caption{\textbf{Entropy maximization time} in the different nanosystems assuming a constant entropy production rate. The larger configurational entropy and reduced entropy production rate obtained in NPMs combine to slow down the thermalization process with respect to the other material systems.}
\label{fig:Fig_05}
\end{figure}


In Fig.~\ref{fig:Fig_05}, we show $\tau_S$ considering supercells consisting of five unit cells, and assuming a constant entropy production rate as characterized in Fig.~\ref{fig:F3}. Accordingly, the modification of the equilibrium phase space and entropy production due to boundaries and nanostructure severely impacts the thermalization process. In particular, the results suggest that realization of NPM nanostructures exhibiting local phonon resonances is a route to profoundly hinder irreversible evolution and delay the emergence of equilibrium, which in turn opens the way for coherent, mechanistic, and precise manipulation of phonons.

\section*{Conclusion}

In conclusion, we introduced a novel framework for the characterization of nanoscale heat conduction phenomena based on quantification of the entropic fingerprint of phonon motion. By equilibrium simulations, we evaluated the influence of nanoscale dynamical constraints on the available phase space for atomic motion, highlighting the additional configurational entropy required to thermalize different nanostructure features when compared to the bulk form. By nonequilibrium simulations, we demonstrated that dynamical constraints induce a degree of reversible evolution in the presence of nanoscale energy currents.~Primarily, our findings reveal that the inclusion of intrinsically resonating nanostructures enables a regime of ordered heat conduction to be approached. This regime is characterized by a strongly fluctuating energy current, where the randomization and thermalization of the system$-$rigorously quantified by the rate of entropy production$-$are hindered and the coherent response of the phonons is distinctly prevalent. Leveraging these effects opens a new horizon for nanoscale phonon engineering with potential implications on a wide range of applications including semiconductor technology, thermoelectricity~\cite{artini2023roadmap}, superconductivity~\cite{lilia2022}, and emerging quantum devices~\cite{giustino2022}$-$all that would benefit from controlled phonon evolution.\\\\ 


\textbf{Author Contributions} \\
{\footnotesize A.B. and M.I.H. conceived, planned, and supervised the research with the stochastic thermodynamics focus led by A.B. and the phonon dynamics focus led by M.I.H. The MD simulations were performed by A.B., P.R. and C.-N.T., and the SED calculations were performed by C.-N.T. All authors participated in the discussion. A.B. and M.I.H. wrote the paper.}\\

\appendix
\section*{Appendix} 
This Appendix covers the geometries and the interatomic potential used in the atomic models (section \ref{potential}) and the methods used to obtain the spectral energy density (SED) spectra (section \ref{sed}), the configurational entropy (section \ref{eq_MD}), and the entropy production (section \ref{neq_MD}), in the different nanostructures from molecular dynamics (MD) simulations using LAMMPS \cite{PLIMPTON19951}. A series of computational verification studies are also included in these sections.

\section{Interatomic potential and detailed geometries}\label{potential}

We use the Stillinger-Weber empirical potential to model the interactions between atoms for all the calculations \cite{SWpotential}. This potential is adequate to reproduce basic properties of bulk silicon such as the thermal conductivity at high temperatures \cite{Schelling2002,Sellan2010}, and has been extensively used in previous studies involving the modeling of silicon nanosystems \cite{Zachariah1996,Honarvar2018,Termentzidis2018}. The methods and approximations used in this work, however, are not restricted to this specific interatomic potential, and can be implemented using refined interatomic potentials, or for the study of other crystalline materials at different temperatures.

In Fig.~\ref{fig:F1}, a periodic unit cell describing the different structures is defined in terms of the number of conventional cells (CC). Each CC is a cube containing 8 atoms with a lattice constant $a=5.431$ \AA. The bulk unit cell has dimensions 6$\times$6$\times$6 CC with periodic boundary conditions applied along all three directions. The membrane unit cell also consists of 6$\times$6$\times$6 CC, but the top and bottom boundaries are free instead of periodic. The NPC unit cell is equivalent to the membrane unit cell, but with a central empty cavity sized 4$\times$4$\times$2 CC. Finally, the NPM unit cell is equivalent to the membrane, but with a nanopillar centered on its top boundary with size 4$\times$4$\times h$ CC, where $h$ denotes the height of the nanopillar. In all cases, the boundary conditions along the in-plane directions prevent drift or global displacements/rotations of the nanostructures.

\section{Spectral energy density (SED)}\label{sed}
We utilize an SED formulation that requires knowledge of only the crystal unit-cell structure and does not require any \it a priori \rm knowledge of the phonon mode eigenvectors. An SED expression, referred to as ${\Phi'}$,  predicts the phonon frequency spectrum~\cite{larkin2014comparison}. As provided in Ref.~\cite{thomas2010predicting}, the SED expression ${\Phi'}$ is a function of wave vector $\pmb{\kappa}$ and frequency $\omega$ and is given by

\begin{equation}
{{{\Phi'}}\left( \pmb{\kappa}, \omega \right)} = \mu_{0} 
\displaystyle\sum\limits_{\alpha}^{3}
\displaystyle\sum\limits_{b}^{n}
\begin{vmatrix}
{\displaystyle\sum\limits_{l}^{N}{ \displaystyle\int\limits_{0}^{\tau_{0}}{\dot{{u}}_{\alpha}
\left (\scriptsize{\!\!\!\!
\begin{array}{l l} 
\begin{array}{l} l\\b
\end{array} \!\!\!\!\!\!
 &;~ t
\!\!\!\!\end{array}}
\right )
e^{{\textrm{i}}\left[\pmb{\kappa}\cdot\pmb{r}_{0}
\left (\!\!
\scriptsize{\begin{array}{l} l\\0
\end{array} \!\!}
\right )-wt\right]} \textrm{d} t}} }
\end{vmatrix}
^{2} ,
\label{eq:SED}
\end{equation}
where $\mu_{0}=m/(4\pi \tau_{0} N)$, $m$ is the mass of a silicon atom, $\tau_{0}$ is the total simulation time, ${\pmb{r}}_{0}$ is the equilibrium position vector of the $l$th unit cell, and ${\dot{{u}}}_{\alpha}$ is the $\alpha$-component of the velocity of the $b$th atom in the $l$th unit cell at time $t$. There are a total of $N=N_{x}\times N_{y}\times N_{z}$ unit cells in the simulated computational domain with $n$ atoms per unit cell. 

We note that in Eq.~(\ref{eq:SED}), the phonon frequencies can be obtained only for the set of allowed wave vectors as determined by the crystal structure. For our model, the $\Gamma X$-path wave vectors are $\kappa_{x}={2\pi j}/({N_{x}6a})$, $j=0$ to $N_{x}/2$. For the computational domain, we set $N_{x} = 50$ and $N_{y}= N_{z}=1$, which gives a $\Gamma X$ wave-vector resolution of ${\Delta}{\kappa}_{x}=0.02 (2\pi/ 6a)$. MD simulations under $NVE$ conditions at 300 K are executed for this system for $2^{22}$ time steps and based on ${\Delta}{t}=0.5$ fs. Equation~(\ref{eq:SED}) is evaluated by computing the Fourier transform of the velocity trajectories extracted every $2^{5}$ steps.

\section{Equilibrium MD and configurational entropy}\label{eq_MD}

We perform equilibrium MD simulations at 300 K to characterize the configurational entropy distributions and the heat flux ACF in the different nanosystems. A single unit cell is simulated for the configurational entropy calculations, and a supercell consisting of five unit cells is considered for the heat flux ACF calculations. The systems are initially equilibrated for 1 ns using a time step 0.5 fs  under the \textit{NPT} ensemble, thus allowing the reconfiguration of the surfaces due to boundary atoms relaxation. 

The excess entropy of each atom is sampled every 6 ps and averaged during the subsequent 6 ns under the \textit{NVE} ensemble. As discussed in \cite{Piaggi2017}, the excess entropy of atom $i$ reads

\begin{equation}\label{excess_entropy}
    s^i=-2\pi\rho k_\textrm{B}\int_0^{r_{m}} [g^i(r)\ln g^i(r)-g^i(r)+1]r^2 \mathrm{d}r,
\end{equation}
where $r$ is the radial distance from atom $i$, $r_m$ is a cut-off distance, $g^i(r)$ is the radial distribution function centered on atom $i$, $\rho$ is the local density around atom $i$, and $k_\textrm{B}$ is the Boltzmann constant. We use the following smoothed distribution $g^i(r)$

\begin{equation}\label{radial_dist}
        g^i(r)=\frac{1}{4\pi\rho r^2}\sum_j \frac{e^{-(r-r_{ij})^2/(2\sigma^2)}}{\sqrt{2\pi\sigma^2}},
\end{equation}
where $r_{ij}$ is the interatomic distance between atom $i$ and its neighbors $j$, and $\sigma$ is the broadening parameter. 

The configurational entropy per atom can be then calculated as $S^{i}=s^{i}+S_{\mathrm{IG}}$, where $S_{\mathrm{IG}}$ is the monoatomic ideal gas entropy per atom as obtained using the Sackur-Tetrode expression

\begin{equation}
    S_{\mathrm{IG}} = k_\textrm{B}\Bigg[ \ln{\Bigg( \frac{V}{N} \Bigg( \frac{4\pi m U_{\mathrm{IG}}}{3 h^2} \Bigg)^{\frac{3}{2}} \Bigg)+ \frac{5}{2}} \Bigg],
\end{equation}
where $N$ is the number of atoms in the system, $V$ is the volume, $U_{\mathrm{IG}}$ is the monoatomic ideal gas energy per atom, and $h$ is the Planck constant.

\subsection{Convergence and size effects}

\begin{figure}[t!] 
\centering 
\includegraphics{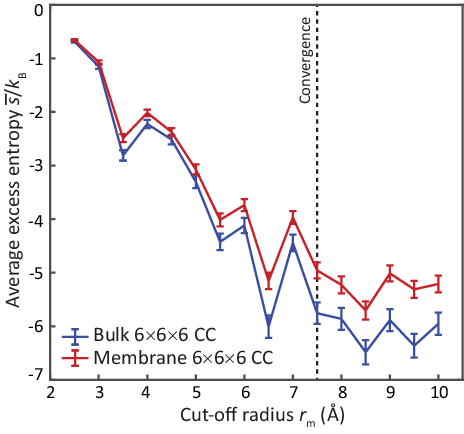}
\caption{\textbf{Cutoff radius convergence.} The excess entropy converges for cutoff radius $r_m>7.5$ \AA. 
The dashed line indicates the value used in the calculations presented in the main article.}
\label{fig:SF1}
\end{figure}

\begin{figure}[t!]
\centering
\includegraphics{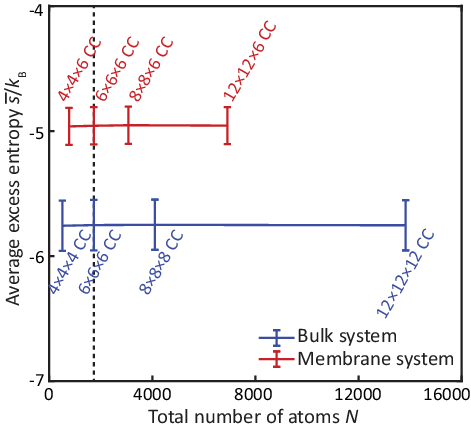}
\caption{\textbf{Computational size effect}. The average excess entropy is independent of the number of atoms considered in the MD simulation for the membrane with a fixed thickness (red) or the bulk (blue) system. The dashed line indicates the unit cell considered in the calculations presented in the main article.}
\label{fig:SF2}
\end{figure}

In this subsection, we investigate the influence of the computational parameters involved in the excess entropy calculations, such as the cut-off radius $r_m$ or the broadening parameter $\sigma$, and we study the relevance of computational size effects in the bulk and membrane systems. 

The radial integration in the $s^i$ formula should include all the neighbors of atom $i$ with a non-negligible contribution. In Fig.~\ref{fig:SF1}, we show that the excess entropy is converged for cutoff radius $r_m\geq 7.5$ \AA. Hence, we use a cutoff radius of $r_m=7.5$ \AA\  for efficient numerical calculation of $s_i$. Moreover, the definition of the radial distribution function in Eq.~\eqref{radial_dist} involves a broadening parameter $\sigma$. This parameter should be chosen small enough to correctly approximate the actual radial distribution function, but large enough to smoothen $g^i(r)$ around the atomic positions and facilitate numerical evaluation. After verifying that a slight modification of this parameter value does not significantly modify the excess entropy integral, we select $\sigma=0.25$ \AA, which is one order of magnitude smaller than the smallest distance between equilibrium atomic positions in the lattice.\\
\indent Finally, as shown in Fig.~\ref{fig:SF2}, we also verified that the average excess entropy in a bulk system (periodic boundary conditions in all directions) or a membrane with a fixed thickness is independent of the size of the unit cell. Therefore, finite computational size effects do not directly impact the configurational entropy values we report. It is also worth emphasizing that the density $\rho$ appearing in Eq.~\eqref{excess_entropy} is the local density around the atom $i$, as required to properly normalize the radial distribution function $g^i(r)$. This is critical for equal-footing comparison of the configurational entropy close to the boundaries and in the internal regions of a given nanosystem. It is observed that if the average global density is used instead, the contribution to the excess entropy at the boundaries is overestimated, which results in a nonphysical low-configurational entropy in some cases. 

\subsection{Internal energy and configurational entropy}

\begin{figure}[t!]
\centering
\includegraphics{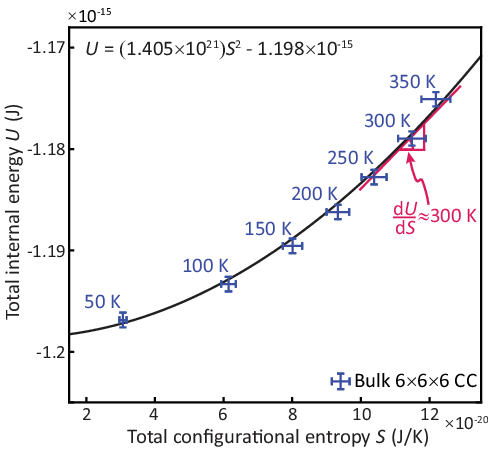}
\caption{\textbf{Total internal energy and total configurational entropy.} The energy versus entropy trend in the bulk system approximately obeys the thermodynamic relation ${\rm d}U/{\rm d}S=T$ around room temperature.}
\label{fig:SF3}
\end{figure}

As described in Section~\ref{Section:configurationalentropy}, the configurational entropy $S$ is estimated by subtracting the excess entropy from the ideal gas entropy. The excess is approximately quantified by considering only two-body correlations in equilibrium MD simulations. We note that computational finite size effects might prevent interpreting $S$ as the equilibrium thermodynamic entropy, since the simulated dynamics might be non-ergodic, meaning that not all the physical microstates that are accessible in equilibrium might be explored with equal probability in the MD calculations. Moreover, it is not evident that alternative methods characterizing the equilibrium entropy in terms of the free energy \cite{Dyre2018} lead to equivalent results. \\
\indent Here we demonstrate the robustness of the present numerical approximations and the physical significance of the resulting entropy by relating it to the internal energy. In Fig.~\ref{fig:SF3}, we show the total internal energy $U$ as a function of the total configurational entropy $S$ in the bulk system at different temperatures $T$, as obtained in equilibrium MD simulations. The obtained trend is approximately obeying the thermodynamic relation ${\rm d}U/{\rm d}S=T$ around room temperature, thus suggesting that the numerical $S$ value correctly approximates the thermodynamic entropy of the considered nanosystem.

\section{Nonequilibrium MD and entropy production} \label{neq_MD}

We perform nonequilibrium MD simulations to compute entropy production lower bounds in each nanosystem. A supercell consisting of five aligned unit cells is simulated in steady state. To prevent rotational or translational effects, a 1.09 nm-thick layer of fixed silicon atoms are added to two opposite ends of each system. Using a Langevin thermostat and a time step of 0.5 fs, the rest of the system is first thermalized at 300 K for 1 ns. Then, a thermal gradient is imposed by thermalizing 2.17 nm-thick layers of atoms in between the fixed ends and the supercell using a Langevin thermostat at 310 K (source) and 290 K (sink), respectively, while leaving the supercell atoms free. The steady state, characterized by a stationary temperature profile between source and sink, is reached by running the simulation for 1 ns with a time step 0.5 fs. Thereafter, stationary evolution is simulated using the same time step during the subsequent 30 ns, wherein the energy transfer between thermostats is evaluated. Specifically, the energy transferred $q_\tau$ corresponds to the total energy introduced in the source to maintain the system out-of-equilibrium in steady state during a time window of $\tau=1$ ns. The values of $\text{E}(q_\tau)$ and $\text{Var}(q_\tau)$ required to calculate the entropy production bound $\widehat{\Sigma}_\tau$ are the mean and the variance of the energy $q_\tau$, respectively. The results are averaged over five simulations with different initial conditions to extract these statistical moments in each case. Similar results are obtained if considering the energy removed from the sink instead of the energy injected in the source to define $q_\tau$.


\subsection{Size effects}

\begin{figure}[t!]
\centering
\includegraphics{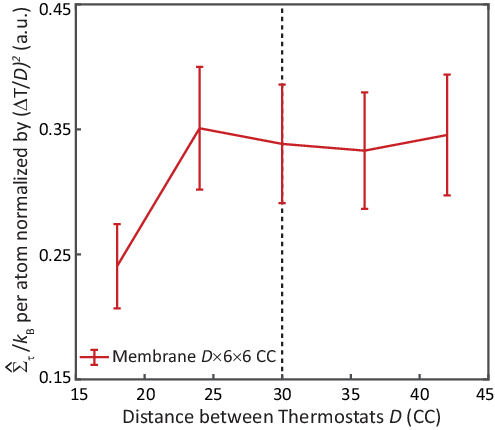}
\caption{\textbf{Computational size effect.} The entropy production is converged for four unit cells of 6$\times$6$\times$6 CC in size between thermostats or larger in the membrane system. Equivalently, convergence is obtained for distances $D$ between thermostats equal to or greater than 24 CC (13.03 nm). The dashed line indicates the value used in the calculations presented in the main article.}
\label{fig:SF4}
\end{figure}

Due to the small size of the simulation domain, incorrect sampling of low frequency phonons, nonphysical coupling between thermostats, and enhanced ballistic phonon transmission can unfold. This would influence the energy transmission within the system \cite{Schelling2002,Sellan2010}, and thus might impact the entropy production characterization. In Fig.~\ref{fig:SF4}, we show $\widehat{\Sigma}_\tau$ as a function of the number of unit cells $D$ between thermostats in the membrane geometry. After adequate normalization by the variable average thermal gradient as per Eq.~\eqref{kappaVSentropy}, we do not observe significant size-dependent entropy production per atom in the membrane for $D>$ 24 CCs. This indicates that the results considering five unit cells between thermostats (i.e., $D=30$ CCs), are not strongly influenced by the small size of the simulation domain. Moreover, we consider the same distance between thermostats for all the nanostructures, and hence the results illustrated in Fig.~\ref{fig:F3} provide an equal-footing comparison among the different nanostructures. We note that $\widehat{\Sigma}_\tau$ is converged for relatively small system sizes $D$ relative to the direct calculation of the thermal conductivity using nonequilibrium simulations \cite{Sellan2010}. We attribute this to the functional form of the entropy bound, where the energy current $q_\tau$ is normalized by combining the mean and the variance. While this significantly reduces the computational cost of the $\widehat{\Sigma}_\tau$ comparison between nanostructures, the present simulation setting might be inadequate to directly compute converged values of the thermal conductivity using the average heat current.\\

\begin{figure}[b!]
\centering
\includegraphics{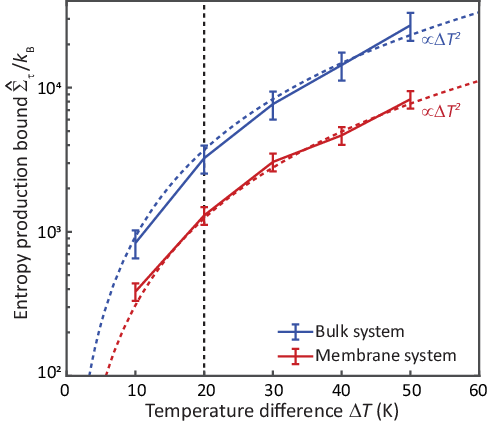}
\caption{\textbf{Influence of the temperature difference between thermostats}. We show the quadratic scaling of the entropy production as a function of $\Delta T$ in the bulk (blue) and the membrane (red). The dashed line indicates the value used in the calculations presented in the main article.}
\label{fig:SF5}
\end{figure}

\begin{figure}[t!]
\centering
\includegraphics{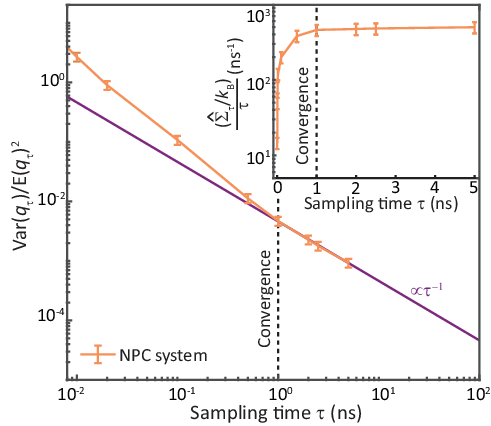}
\caption{\textbf{Influence of sampling time on entropy production rate.} In the NPC system, we show that the statistical moments of the energy current obey the long sampling time limit according to the central limit theorem for $\tau\geq1$ ns. The inset illustrates that the entropy production rate bound $\widehat{\Sigma}_\tau/\tau$ consequently converges for $\tau\geq$1 ns. The dashed line indicates the value used in the calculations presented in the main article.}
\label{fig:SF6}
\end{figure}

\begin{figure}[b!]
\centering
\includegraphics{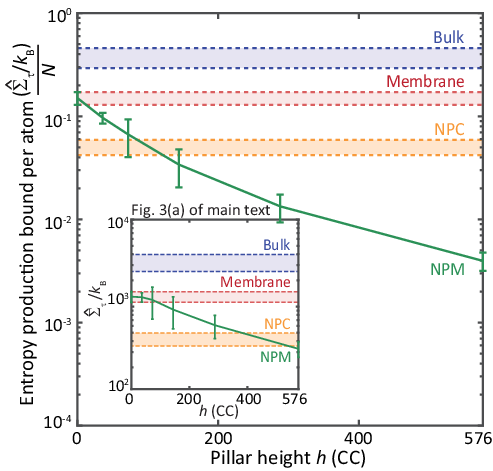}
\caption{\textbf{Entropy production normalized by the number of atoms.} As a reference, the inset shows the total entropy production.}
\label{fig:SF7}
\end{figure}
\indent Moreover, in the NPM structure, we consider small inter-nanopillar distance compared to the nanopillar height to explore the effects of the phonon resonances in extreme conditions. Consequently, the nanopillars from different unit cells may bend during the simulation sufficiently enough to touch each other. In realistic NPM samples, however, the ratio of the distance between the nanopillars to the nanopillar height will necessarily be larger, and contact between nanopillars is not expected. To mimic realistic conditions in larger NPMs, we prevent inter-nanopillar contact in our simulations by implementing a weak Lennard-Jones repulsion between atoms from different nanopillars~\cite{Tsai2023PhD}. For consistency, the heat flux ACF calculations in Fig.~\ref{fig:Fig_04} also consider this weak repulsion between contiguous nanopillars. This strategy maintains the crystalline order of the lattice in NPM models even for relatively tall nanopillars, and has a small effect on the phonon dispersion relations. 

Nonetheless, for $h>72$ CCs, we observe a slight decrease in the potential energy in the nanopillars due to nanopillar bending and interactions after the thermostats are connected to the system. This loss of potential energy translates to a small temperature rise in the nanopillars and delays the emergence of the steady-state conditions, where the average energy injection rate in the hot thermostat is equivalent to the removal rate in the cold one. To test the influence of these effects on $\widehat{\Sigma}$ in the NPM system, we performed an additional set of simulations considering an extra thermalization step before characterizing $\widehat{\Sigma}_\tau$. First, we verify that the nanopillar potential energy stabilizes under the influence of the thermostats. Then, we remove all the residual kinetic energy by disconnecting the thermostats and thermalizing all the system at 300 K for 1 ns. Finally, the steady-state conditions are established again for 1 ns before characterization of the statistical moments of $q_\tau$ during the subsequent 30 ns. In addition, we discard a few simulation realizations where excessive bending of the nanopillars is identified. This simulation protocol resulted in the same trends of the flux fluctuations and $\widehat{\Sigma}$ as a function of the nanopillar height reported in Fig.~\ref{fig:F3}, thus confirming that inter-nanopillar contact does not have a dominant effect on the present results. However, the additional thermalization step reduces $\text{Var}(q_\tau)$ and increases $\bar{\Sigma}$ for very tall nanopillars ($h>$144 CC).

\subsection{Influence of temperature difference between thermostats}

We consider a temperature difference of $\Delta T=20$ K between the thermostats in the MD simulations to estimate $\widehat{\Sigma}_\tau$ in Fig.~\ref{fig:F3}. By increasing $\Delta T$, we observe a quadratic increase in the entropy production for both the bulk and the membrane systems, as shown in Fig.~\ref{fig:SF5}. For larger temperature differences, the system is perturbed further away from equilibrium, and a proportionally larger flow of energy is established in the system. Since the fluctuations are seen not to increase quadratically with respect to the temperature difference, the system can produce more entropy and irreversibly dissipate more energy when the temperature difference between the thermostats is larger.

The quadratic scaling $\widehat{\Sigma}_\tau\propto \Delta T^2$ can be interpreted using the relation~\eqref{kappaVSentropy} between the thermal conductivity $k$ and the entropy production \cite{Klemens1989}.
Accordingly, since the thermal conductivity is an intrinsic system property independent of the imposed temperature gradient $\nabla{T}=\Delta T/D$, the entropy production scales as the square of $\Delta T$ for a fixed system size $D$, consistent with the results shown in Fig.~\ref{fig:SF5}. Finally, this geometry-independent scaling justifies the arbitrary choice $\Delta T=20$ K used in this work to compare the effects of nanostructuring on the entropy production using MD simulations.

\subsection{Sampling time and central limit theorem}\label{centrallimittheorem}

In the limit of long sampling times $\tau$, the central limit theorem stipulates that $\text{Var}(q_\tau)/\text{E}(q_\tau)^2\sim 1/\tau$ \cite{Song2021}. In Fig.~\ref{fig:SF6}, we verify that our statistical analysis of the nonequilibrium MD simulations is consistent with this theorem for $\tau\geq$ 1 ns.\\
\indent Motivated by this result, we use a sampling time $\tau=1$ ns for our estimations of the production of entropy. This corresponds to the minimum sampling time $\tau$ that allows us to characterize an entropy production rate $\widehat{\Sigma}_\tau/\tau$ independent of $\tau$ (see inset of Fig.~\ref{fig:SF6}). Nevertheless, we emphasize that, even for smaller sampling times, the relative comparison of the entropy production between the different nanostructures remains the same.

\subsection{Normalized entropy production}

The entropy maximization and thermalization in our nanosystem models are achieved via anharmonic phonon-phonon interactions. Hence, the entropy production of a nanosystem is expected to decrease by reducing its size to scales smaller than the phonon mean free paths and scattering times. In fact, we obtain very strong reductions in the production of entropy in the nanostructured geometries relative to the bulk. This trend explains why, in general, nonequilibrium thermodynamic behavior tends to persist in semiconductors at reduced length or time scales. \\
\indent Nevertheless, we note that in the NPM geometry, reducing the size of the system by considering smaller nanopillars results in an increase in entropy production rather than a decrease. This emphasizes the role of the phonon resonances and suggests that the emerging phonon coherence at the nanoscale influences entropic evolution in unexpected ways. The complex connections between mechanistic phonon behavior and the global thermodynamic response are thus manifested in a system such as the NPM. For illustration, in Fig.~\ref{fig:SF7}, we show the entropy production normalized by the total number of atoms. Contrary to what would be expected for standard structures, larger NPM systems produce less entropy and thus relax slower.

\bibliographystyle{RS}
\bibliography{biblio}

\end{document}